\documentclass[prb,twocolumn,showpacs,graphics,preprintnumbers]{revtex4}
\usepackage{graphicx}
\usepackage{amsmath}
\begin{document}
\preprint{}

\title{Lower Critical Fields of Superconducting PrFeAsO$_{1-y}$ Single Crystals}

\author{R.~Okazaki,$^\mathrm{1}$ M.~Konczykowski,$^\mathrm{2}$ 
C.~J.~van~der~Beek,$^\mathrm{2}$ T.~Kato,$^\mathrm{1}$
K.~Hashimoto,$^\mathrm{1}$ M.~Shimozawa,$^\mathrm{1}$ 
H.~Shishido,$^\mathrm{1}$ M.~Yamashita,$^\mathrm{1}$ 
M.~Ishikado,$^\mathrm{3}$ H.~Kito,$^\mathrm{4,5}$ A.~Iyo,$^\mathrm{4,5}$ 
H.~Eisaki,$^\mathrm{4,5}$ S.~Shamoto,$^\mathrm{3,5}$ 
T.~Shibauchi,$^\mathrm{1}$ and Y.~Matsuda$^\mathrm{1,2}$}

\affiliation{$^1$Department of Physics, Kyoto University, Kyoto 
606-8502, Japan }%
\affiliation{$^2$Laboratorie des Solides Irradi\'es, CNRS-UMR 7642 \& CEA/DSM/IRAMIS, Ecole Polytechnique, 91128, Palaiseau, France}%
\affiliation{$^3$Quantum Beam Science Directorate, Japan Atomic 
Energy Agency, Tokai, Naka, Ibaraki 319-1195, Japan}%
\affiliation{$^4$Nanoelectronics Research Institute (NeRI), National 
Institute of Advanced Industrial Science and Technology (AIST), 1-1-1 
Central 2, Umezono, Tsukuba, Ibaraki 305-8568, Japan}%
\affiliation{$^5$JST, TRIP, Chiyoda, Tokyo 102-0075, Japan}%

\begin{abstract}

We have studied the lower critical fields $H_{c1}$ of superconducting 
iron oxipnictide PrFeAsO$_{1-y}$ single crystals for {\boldmath $H$} 
parallel and perpendicular to the $ab$-planes.  Measurements of the 
local magnetic induction at positions straddling the sample edge by 
using a miniature Hall-sensor array clearly resolve the first flux 
penetration from the Meissner state.   The temperature dependence of 
$H_{c1}$ for {\boldmath $H$}$\parallel c$ is well scaled by the 
in-plane penetration depth without showing any unusual behavior, in 
contrast to previous reports.  The anisotropy of penetration lengths 
at low temperatures is estimated to be $\simeq$ 2.5, which is  considerably 
smaller than the anisotropy of the coherence lengths.  This is indicative of 
multiband superconductivity in this system, in which the active band 
for superconductivity is more anisotropic.  We also point out 
that the local induction measured at a position near the center 
of the crystal, which has been used in a number of reports for the 
determination of $H_{c1}$, might seriously overestimate the obtained
$H_{c1}$-value. 

\end{abstract}


\maketitle

\section{Introduction}

The recent discovery of high temperature superconductivity in 
Fe-based compounds has attracted considerable interest.\cite{Kam08}  
In this new class of compounds with a very low carrier 
density,\cite{Tak08,Che08Ce,Ren08Pr,Kit08Nd,Ren08Nd,Che08Sm,Yan08Gd,Wan08Gd} 
superconductivity occurs in  proximity to a magnetic instability, and 
unconventional pairing mechanisms mediated by magnetic fluctuations 
have been proposed by several groups.\cite{Maz08,Kur08,Aok08}
One of the remarkable features, which is in sharp contrast to the 
high-$T_c$ cuprates,  appears to be the multiband nature of  
superconductivity, in electron and hole pockets.\cite{Sin08}   Recently, a 
multiband effect on superconductivity has been reported in 
several compounds.\cite{Sey05,Kas07,Nak08}  In
particular,  the two-gap superconductivity in MgB$_2$ manifests 
itself in the unusual temperature- and magnetic field dependence of 
the anisotropy parameters in the superconducting 
state.\cite{Bou02,Lya04,Fle05} 
However, the crucial difference is that the interband coupling is 
very weak in MgB$_2$, while in Fe-based compounds nesting between the 
hole- and electron bands was suggested to be important for the 
occurrence of high temperature superconductivity.\cite{Maz08,Kur08,Aok08,Ike08,Nom08,Seo08,Tes08_1}
In this context, a detailed clarification of the multiband nature of 
superconductivity in the Fe-based oxypnictides is indispensable for 
the elucidation of the superconducting properties, and especially for 
the pairing mechanism.

An accurate determination of the lower critical field $H_{c1}$ is an 
important means to clarify not only the superconducting gap 
symmetry, but also the multiband nature of  superconductivity.  
However, the reliable measurement of the lower critical field is a 
difficult task, in particular when strong vortex pinning 
is present.  
We also point out that to date the reported values of anisotropy parameter strongly vary\cite{Mar08,Wey08,Bal08,Kub08} spanning from 1.2 (Ref.~\onlinecite{Kub08}) up to $\sim 20$ (Ref.~\onlinecite{Wey08}), which may be partly due to the effects of strong pinning. 
In this study,  we use an 
unambiguous method to avoid this  difficulty associated with pinning, 
by determining $H_{c1}$ as the field $H_p$ at which first flux penetration occurs 
from the edge of the crystal. This allows us to extract the temperature dependent values of the lower critical fields parallel 
to the $c$-axis ($H_{c1}^c$) and the $ab$-plane ($H_{c1}^{ab}$), respectively
as well as the anisotropy parameter $H_{c1}^c/H_{c1}^{ab}$ in single crystals of Fe-based 
superconductors. 

We  directly determine $H_{p}$ by measuring the magnetic
induction just inside and outside the edge of the single 
crystals, by using a miniature Hall-sensor array.   
First, we show that local magnetization measurements at a 
position near the center of the crystal, which have been  
used by several groups for the determination of $H_{c1}$,  seriously 
overestimate $H_{c1}$ in systems with strong pinning.   
Second, we find that the temperature dependence of $H_{c1}$ 
determined at the edge does not show any unusual 
behavior\cite{Ren08_1111} and is well scaled by the penetration depth 
results measured on the crystal in the same batch.\cite{has08}   
Finally, we find that the anisotropy of the penetration depths  
$\gamma_{\lambda}\equiv \lambda_c/\lambda_{ab} \simeq 
H_{c1}^c/H_{c1}^{ab}$, where $\lambda_c$ and $\lambda_{ab}$ are 
out-of-plane and in-plane penetration depths, respectively, is much 
smaller than the anisotropy of the coherence lengths 
$\gamma_{\xi}\equiv \xi_{ab}/\xi_c= H_{c2}^{ab}/H_{c2}^c$, where 
$\xi_{ab}$ and $\xi_c$ are in- and out-of- plane coherence lengths, 
respectively, and $H_{c2}^{ab}$ and $H_{c2}^c$ are the upper critical 
fields parallel and perpendicular to the $ab$-plane, respectively.  
This result provides strong evidence for the multiband nature of the 
superconductivity.

\section{Experimental}

\begin{figure}[t]
\begin{center}
\includegraphics[width=7cm]{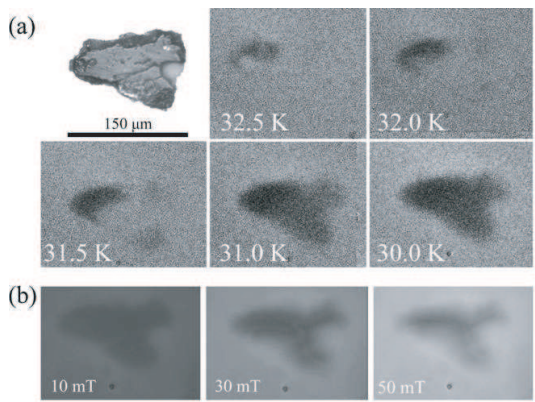}
\caption{(color online). (a) Differential magneto-optics images of PrFeAsO$_{1-y}$ 
$\#$1 with field modulation $\delta B =$~0.2~mT in zero field. The 
Meissner screening occurs completely within narrow temperature range. 
(b) MO images at $T$~=~7.1~K. Magnetic flux penetrates from the edge 
of the crystal and the field distribution shows the Bean critical 
state.}
\label{mo}
\end{center}
\end{figure}

Experiments have been performed on high-quality PrFeAsO$_{1-y}$ 
single crystals, grown by a high-pressure synthesis method 
using a belt-type anvil apparatus (Riken CAP-07). Powders of PrAs, 
Fe, Fe$_2$O$_3$ were used as the starting materials. PrAs was 
obtained by reacting Pr chips and As pieces at 500$^{\circ}$C for 10 
hours, followed by a treatment at 850$^{\circ}$C for 5 hours in an 
evacuated quartz tube. 
The starting materials were mixed at nominal compositions of 
PrFeAsO$_{0.6}$ and ground in an agate mortar in a glove box filled 
with dry nitrogen gas. The mixed powders were pressed into pellets. The 
samples were then grown by heating the pellets in BN crucibles under a 
pressure of about 2~GPa at 1300$^{\circ}$C for 2 hours.  
Platelet-like single crystals of dimensions up $150 \times 150 \times 30$~$\mu$m$^3$ 
were mechanically selected from the polycrystalline pellets.   
The single crystalline nature of the samples was checked by Laue X-ray diffraction.\cite{Has_JPSJ} 
Our crystals, whose $T_c$ $(\approx34$~K) is lower than the optimum 
$T_c \approx$ 51~K of PrFeAsO$_{1-y}$,\cite{Ren08_PD} are in the 
underdoped regime ($y\sim0.1$),\cite{Lee08} which is close to the spin-density-wave 
order.\cite{Zha08} 
The sample homogeneity was checked by magneto-optical (MO) 
imaging.  MO images of PrFeAsO$_{1-y}$ sample $\#$1 ($\sim 135\times 63 
\times 18$ $\mu$m$^3$) are shown in Fig.~\ref{mo}(a). 
The crystal exhibits a nearly perfect Meissner state $\sim$~2~K below 
$T_c$; no weak links are observed, indicating a good homogeneity.  
At low temperatures, the magnetic field distribution is well 
described by the Bean critical state model as shown in 
Fig.~\ref{mo}(b).\cite{Zeldov94}  

The local induction near the surface of the platelet crystal has been measured by 
placing the sample on top of a miniature Hall-sensor array tailored 
in a GaAs/AlGaAs 
heterostructure.\cite{Shi07}
Each Hall sensor has an active area of $3 \times 3$ $\mu$m$^2$; the
center-to-center distance of neighboring sensors is 20~$\mu$m.  The 
local induction at the edge of the crystal was detected by the 
miniature Hall sensor located at $\leq 10$~$\mu$m from the edge.  
The magnetic field $H_a$ is applied for {\boldmath $H$}$\parallel c$ 
and {\boldmath $H$}$\parallel ab$-plane by using a low-inductance 
2.4~T superconducting magnet with a negligibly small remanent field.  

The in-plane resistivity is measured by the standard four-probe 
method under magnetic fields up to 10 T. 
The electrical contacts were attached by using the W deposition 
technique in a Focused-Ion-Beam system.

\section{Results and discussion}

\begin{figure}[t]
\begin{center}
\includegraphics[width=8.5cm]{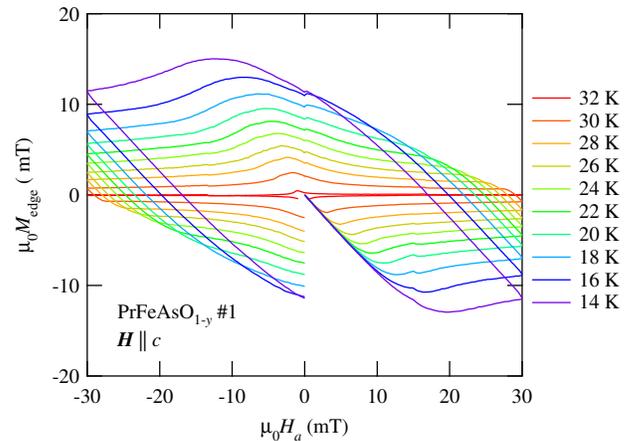}
\caption{(color online). Local magnetization loops for {\boldmath 
$H$}$\parallel c$, measured by the miniature Hall sensor located at 
$\leq$~10~$\mu$m from the edge of the crystal.}
\label{MH}
\end{center}
\end{figure}

In Fig.~\ref{MH} we show the field dependence of the ``local 
magnetization'', $M_{\rm edge} \equiv \mu_0^{-1}B_{\rm edge} - H_a$, at the edge of the 
crystal, for {\boldmath $H$}$\parallel c$, measured after zero field 
cooling.  After the initial negative slope corresponding to the Meissner state, vortices enter 
the sample and $M_{\rm edge}(H_{a})$ shows a large hysteresis. 
The shape of the magnetization loops (almost symmetric about the 
horizontal axis) indicates that the hysteresis mainly arises from  
bulk flux pinning rather than from the (Bean-Livingston)  surface 
barrier.\cite{Kon99}

\begin{figure}[t]
\begin{center}
\includegraphics[width=8.5cm]{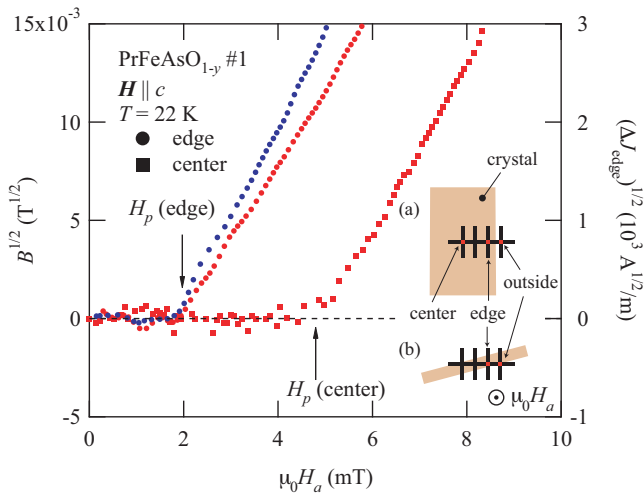}
\caption{(color online). Typical curves of $\sqrt{B}$ 
(left axis) at the edge (circles) and at the center (squares) of the 
crystal and $\sqrt{\Delta j_{\rm edge}}$ (right axis) plotted as a 
function of $H_a$ for {\boldmath $H$}$\parallel c$ at $T$~=~22~K, in 
which $H_a$ is increased after ZFC. The insets are schematic 
illustrations of the experimental setup for {\boldmath $H$}$\parallel 
c$ (a) and {\boldmath $H$}$\parallel ab$-plane (b).}
\label{DM}
\end{center}
\end{figure}

As shown in Fig.~\ref{MH}, the initial slope of the magnetization 
exhibits a nearly perfect linear dependence, $M_{\rm edge}=-\alpha 
H_a$.  Since the Hall sensor is placed on the top surface, with a small 
but non-vanishing  distance between the sensor and the crystal, the magnetic 
field leaks around the sample edge with the result that the slope $\alpha$ is slightly smaller than unity. 
Figure~\ref{DM} shows typical curves of $B^{1/2} \equiv \mu_{0}^{1/2}(M + \alpha 
H_a)^{1/2}$ at the edge (circles) and at the 
center (squares) of the crystal, plotted as a function of $H_a$; the 
external field orientation {\boldmath $H$}$\parallel c$ and $T$~=~22~K.  The $\alpha H_a$-term is obtained by 
a least squares fit of the low-field magnetization.  The first penetration 
field $H_p$ corresponds to the field $H_{p}$(edge), above which 
$B^{1/2}$ increases almost linearly,  is clearly 
resolved. In Fig.~\ref{DM}, we show the equivalent curve, measured at the center of the 
crystal.  At the center, $B^{1/2}$ also increases 
linearly,  starting from a larger field, $H_p$(center).

We have measured the positional dependence of $H_p$ and observed that 
it increases with increasing distance from the edge.  To 
examine whether $H_p$(edge), \em i.e. \rm $H_{p}$ measured at $\leq$~10~$\mu$m from the 
edge, truly corresponds to the field of first flux penetration at the boundary of the 
crystal, we have determined the local screening current density $j_{\rm 
edge} = \mu_0^{-1} (B_{\rm edge}-B_{\rm outside})/\Delta x$ at the 
crystal boundary. Here $B_{\rm edge}$ is the local magnetic induction measured 
by the sensor just inside the edge, and $B_{\rm outside}$ is the 
induction measured by the neighboring sensor just outside the edge.
 For fields less than the first penetration field, $j_{\rm edge} \simeq \beta H_{a}$
 is the Meissner current, which is simply proportional to the 
applied field ($\beta$ is a constant determined by geometry). At $H_{p}$, the screening 
current starts to deviate from linearity. Figure~\ref{DM} shows the 
deviation $\Delta j_{\rm edge} \equiv j_{\rm edge}-\beta H_a$ as a function of $H_a$. 
 As depicted in Fig.~\ref{DM},  $\sqrt{\Delta j_{\rm edge}}$ again increases 
linearly with $H_a$ above $H_p$(edge).  This indicates that the 
$H_p$(edge) is very close to the true field of first flux penetration.  

\begin{figure}[t]
\begin{center}
\includegraphics[width=8cm]{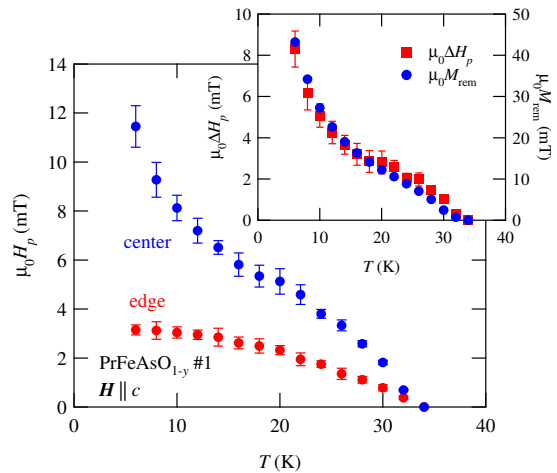}
\caption{(color online). The temperature dependence of the flux 
penetration fields $H_p$ at the edge and the center of the crystal. 
The inset shows the temperature dependence of the difference between 
$H_p$ in the center and at the edge (left axis), as well as the remanent 
magnetization $M_{\rm rem}$ (right axis).}
\label{center_edge}
\end{center}
\end{figure}

In Fig.~\ref{center_edge}, we compare the temperature dependence of 
$H_{p}$(edge) and $H_p$(center).  In the whole temperature range, $H_p$(center) well exceeds 
$H_p$(edge).  Moreover,   $H_p$(center)  increases with decreasing $T$ without any 
tendency towards saturation.  In sharp contrast, $H_p$(edge) saturates at low 
temperatures.   The inset of Fig.~\ref{center_edge} shows the difference 
between $H_p$ measured in the center and at the edge, $\Delta H_p=H_p$(center) $-$ $H_p$(edge).  
$\Delta H_p$ increases steeply with decreasing temperature.  
Also plotted in the inset of Fig.~\ref{center_edge} is the remanent 
magnetization $M_{\rm rem}$ (\em i.e. \rm the $H_{a} = 0$ value of 
$M_{\rm edge}$ on the decreasing field branch), measured at near the 
crystal center. This is proportional to the critical current 
density $j_c$ arising from flux pinning. 
The temperature dependence of $\Delta H_p$ is very similar to that of 
$j_{c}$, which indicates that $H_p$(center) is strongly influenced by pinning.   
Hence, the present results demonstrate that the lower critical field 
value determined by local magnetization measurements carried out at positions 
close to the crystal center, such as reported by several groups,  is 
affected by vortex pinning effects and might be seriously 
overestimated.\cite{Ren08_1111,Ren08_122}

\begin{figure}[t]
\begin{center}
\includegraphics[width=8.5cm]{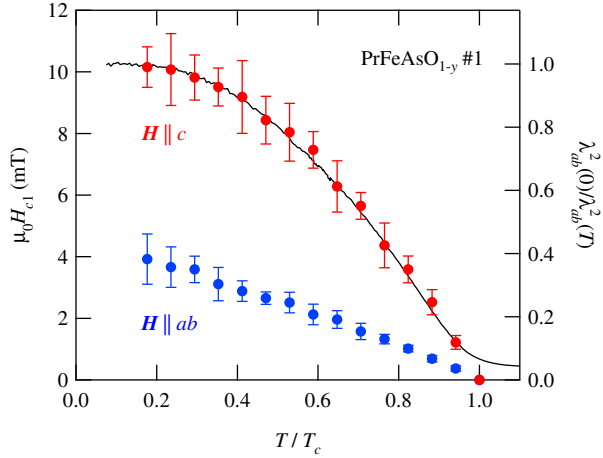}
\caption{(color online). Lower critical fields as a function of 
temperature in PrFeAsO$_{1-y}$ single crystals (left axis). The solid 
line (right axis) presents the superfluid density 
$\lambda^2_{ab}(0)/\lambda^2_{ab}(T)$ determined by  surface 
impedance measurements on crystals from the same 
batch.\cite{has08}}
\label{Hc1}
\end{center}
\end{figure}

The absolute value of $H_{c1}$ is evaluated by taking into account 
the demagnetizing effect.  For a platelet sample, $H_{c1}$ is given by  
\begin{equation}
H_{c1}=H_p/\tanh \sqrt{0.36b/a}
\label{Brandt}
\end{equation}
where $a$ and $b$ are the width and the thickness of the crystal, 
respectively.\cite{Bra99} In the situation where {\boldmath $H$}$\parallel 
c$, $a=$ 63 $\mu$m and $b=$ 18 $\mu$m, while  $a=$ 18 $\mu$m and $b=$ 63 $\mu$m for {\boldmath $H$}$\parallel 
ab$-plane. These values yield  $H_{c1}^c = 3.22 H_p$ and $H_{c1}^{ab} = 1.24 H_p$, 
respectively. In Fig.~\ref{Hc1}, we plot $H_{c1}$ as a function of temperature both 
for {\boldmath $H$} $\parallel c$ and {\boldmath $H$} $\parallel 
ab$-plane. The solid line in Fig.~\ref{Hc1} indicates the temperature dependence 
of the superfluid density normalized by the value at $T=0$~K, which is
obtained from $ab$-plane penetration depth measurements of a 
sample from the same batch.\cite{has08}
 $H_{c1}^c(T)$ is well scaled by the superfluid density, which is 
consistent with fully gapped superconductivity; it does not show the 
unusual behavior reported in Ref.~\onlinecite{Ren08_1111}.
To roughly estimate the in-plane penetration depth at low 
temperatures, we use the approximate single-band London formula,
\begin{equation}
\mu_0H_{c1}^{c} = \frac{\Phi_0}{4\pi \lambda_{ab}^2}\left[ 
\ln\frac{\lambda_{ab}}{\xi_{ab}}+0.5 \right]\\
\end{equation}
where $\Phi_0$ is the flux quantum.   
Using $\ln\lambda_{ab}/\xi_{ab} + 0.5 \sim 5$, we obtain 
$\lambda_{ab} \sim$~280~nm.
This value is in close correspondence with the $\mu$SR results in slightly underdoped LaFeAs(O,F).\cite{Lut08}

\begin{figure}[t]
\begin{center}
\includegraphics[width=8.5cm]{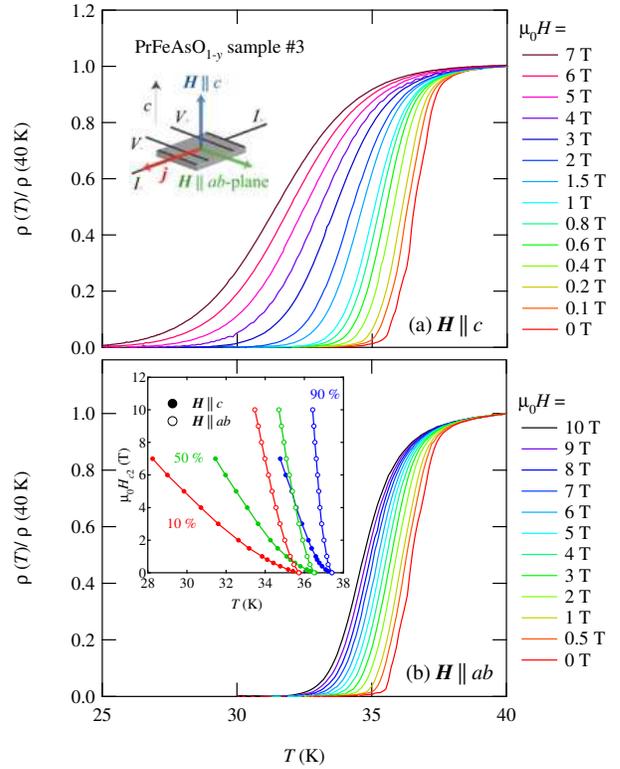}
\caption{(color online). Temperature dependence of the in-plane 
resistivity in PrFeAsO$_{1-y}$ single crystals for {\boldmath 
$H$}$\parallel c$ (a) and {\boldmath $H$}$\parallel ab$-plane (b). 
Inset shows the temperature dependence of the upper critical fields 
$H_{c2}$ determined by several criteria that the resistivity reaches 
10$\%$, 50$\%$, and 90$\%$ of the normal-state resistivity. The 
experimental configuration is also sketched.}
\label{rho}
\end{center}
\end{figure}

Figures~\ref{rho}(a) and (b) depict the temperature dependence of the 
in-plane resistivity for {\boldmath $H$}$\parallel c$ and {\boldmath 
$H$}$\parallel ab$-plane, respectively.  
In the inset of Fig.~\ref{rho}(b), we display the 
fields at which the resistivity is equal to 10$\%$, 50$\%$, and 90$\%$ of 
the normal-state resistivity. For sufficiently high magnetic field, 
these resistance loci are roughly proportional to the upper critical field.
In zero field, the resistive transition exhibits a rather sharp 
transition with the transition width $\Delta T_c \approx$~2~K.  
By applying a magnetic field along the $c$-axis, the transition shifts 
 to slightly lower temperatures and becomes broadened. The 
resistive transition curves broaden less for {\boldmath 
$H$}$\parallel ab$-plane.  
These results indicate that the anisotropy of the upper critical 
fields in the present system is rather large and that fluctuation effects 
play an important role for the transition in magnetic fields,\cite{URu} similar 
to high-$T_c$ cuprates.\cite{Kwok}

\begin{figure}[t]
\begin{center}
\includegraphics[width=7.5cm]{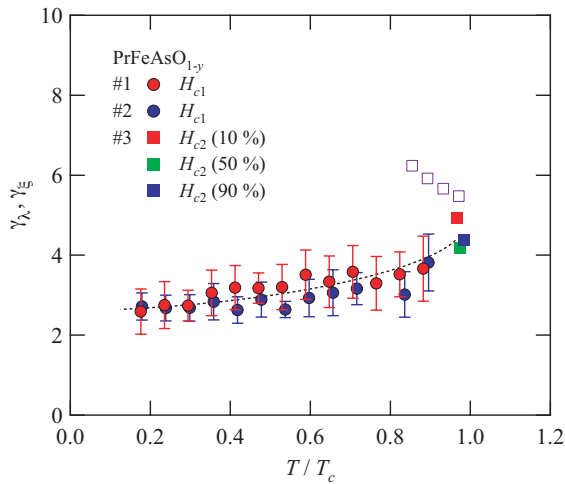}
\caption{(color online). Normalized temperature dependence of the 
anisotropies of $H_{c1}$ ($\gamma_{\lambda}$, closed circles) and $H_{c2}$ ($\gamma_{\xi}$, closed squares) 
in PrFeAsO$_{1-y}$ single crystals. The anisotropy of $H_{c2}$ in 
NdFeAsO$_{0.82}$F$_{0.18}$ ($\gamma_{\xi}$, open squares) measured by Y. Jia {\it et 
al.}\cite{Jia08Nd} is also plotted. The dashed line is a guide to the eye.}
\label{anisotropy}
\end{center}
\end{figure}

Finally, Fig.~\ref{anisotropy} shows the anisotropy of the lower 
critical fields, $\gamma_{\lambda}$ obtained from the results in Fig.~\ref{Hc1}.  Here, since the penetration lengths are much larger than the coherence lengths for both {\boldmath $H$}$\parallel ab$ and   {\boldmath $H$}$\parallel c$, the logarithmic term in Eq.(2) does not strongly depend on the direction of magnetic field.  We thus assumed  $H_{c1}^c/H_{c1}^{ab}\simeq \lambda_c/\lambda_{ab}$.  
The anisotropy $\gamma_{\lambda}\approx 2.5$ at very low temperature, and increases 
gradually with temperature.  In Fig.~\ref{anisotropy}, the anisotropy 
of the upper critical fields $\gamma_{\xi}$ is also plotted, where 
$\gamma_{\xi}$ is determined by the loci of 10\%, 50\% and 90\% of 
the normal-state resistivity  (see the inset of Fig.~\ref{rho}(b)).  
Since $H_{c2}$ increases rapidly and well exceeds 10~T just below 
$T_c$ for {\boldmath $H$}$\parallel ab$,  plotting $\gamma_{\xi}$ is restricted to a narrow temperature 
interval. In Fig.~\ref{anisotropy}, we also plot the 
$H_{c2}$-anisotropy data measured on NdFeAsO$_{0.82}$F$_{0.18}$ by 
the authors of  Ref.~\onlinecite{Jia08Nd}. These indicate that the temperature dependence 
of $\gamma_{\lambda}$ is markedly different from that of 
$\gamma_{\xi}$.  

According to the anisotropic Ginzburg-Landau (GL) equation in single-band superconductors,   $\gamma_{\lambda}$ should coincide with  
$\gamma_{\xi}$ over the whole temperature range.  
Therefore, the large difference between these anisotropies provides 
strong evidence for multiband superconductivity in the present 
system.  We discuss the anisotropy parameters for the multiband 
superconductivity below.  
According to GL theory, $\gamma_{\lambda}$ and $\gamma_{\xi}$ at 
$T_c$ are given as 
\begin{equation}
\gamma_{\xi}^2(T_c)=\gamma_{\lambda}^2(T_c)=\frac{\langle \Omega^2 
v_a^2 \rangle}{\langle \Omega^2 v_c^2 \rangle},
\end{equation}
where $\langle \cdot \cdot \cdot \rangle$ denotes the average over 
the Fermi surface, $v_a$ and $v_c$ are the Fermi velocities parallel 
and perpendicular to the $ab$-plane, respectively.\cite{Kog02,Mir03}  
$\Omega$ represents the gap 
anisotropy ($\langle\Omega^2\rangle = 1$), which is related to the pair potential $V(${\boldmath 
$v,v'$}$)=V_0\Omega(${\boldmath $v$})$\Omega(${\boldmath $v'$}).  At 
$T=0$~K, the anisotropy of the penetration depths is
\begin{equation}
\gamma_{\lambda}^2(0)=\frac{\langle v_a^2 \rangle}{\langle v_c^2 
\rangle}.
\end{equation}
The gap anisotropy does not enter $\gamma_{\lambda}(0)$, while 
$\gamma_{\xi}$ at $T=0$~K is mainly determined by the gap anisotropy of 
the active  band responsible for superconductivity.  Thus the gradual reduction of $\gamma_{\lambda}$ with decreasing temperature can be accounted for by considering that the contribution of the gap anisotropy diminished at low temperatures.  This also implies that the superfluid density along the $c$-axis $\lambda_c^2(0)/\lambda_c^2(T)$ has steeper temperature dependence than that in the plane $\lambda_{ab}^2(0)/\lambda_{ab}^2(T)$.
A pronounced discrepancy between $\gamma_{\xi}$ and 
$\gamma_{\lambda}$ provides strong evidence for the multiband 
nature of superconductivity in PrFeAsO$_{1-y}$, with different 
gap values in different bands.
We note that similar differences between $\gamma_{\xi}(T)$ and $\gamma_{\lambda}(T)$, as well as $\lambda_c^2(0)/\lambda_c^2(T)$ and $\lambda_{ab}^2(0)/\lambda_{ab}^2(T)$, have been reported in the two-gap 
superconductor MgB$_2$.\cite{Lya04,Fle05}
We also note that angle-resolved photoemission spectroscopy 
(ARPES),\cite{Din08} Andreev reflection,\cite{Sza08} and penetration depth\cite{Has122} measurements on (K$_{1-x}$Ba$_x$)Fe$_2$As$_2$ and NMR\cite{Mat08} and penetration depth\cite{Mal08} studies of LnFeAs(O,F) (Ln = Pr, Sm) have suggested 
 multiband superconductivity with two gap values in Fe-based 
oxypnictides.

Band structure calculations for LaFeAsO$_{1-x}$F$_x$ yield an 
anisotropy of the resistivity of approximately 15 for isotropic 
scattering,\cite{Sin08} which corresponds to $\gamma_{\lambda}\sim 
4$. This value is close to the observed value. 
The fact that $\gamma_{\xi}$ well exceeds $\gamma_{\lambda}$ 
indicates that the active band for superconductivity is more 
anisotropic than the passive band. According to band structure calculations, 
there are five relevant bands in  LaFeAsO$_{1-x}$F$_x$. Among them, one of the three hole bands near the 
$\Gamma$ point and the electron bands near the M point are two-dimensional and cylindrical.
 The other two hole bands near the $\Gamma$ point have more dispersion 
along the $c$ axis,\cite{Sin08} although the shape of these Fermi surfaces is 
sensitive to the position of the As atom with respect to the Fe plane,
which in turn depends on the rare earth.\cite{Vil08}
Our results implying that the active band is more anisotropic 
is in good correspondence with the view that the nesting between the 
cylindrical hole and electron Fermi surfaces is essential for 
superconductivity. This is expected to make these two-dimensional bands the active 
ones, with a large gap, and the other more three-dimensional bands passive ones with 
smaller gaps.

\section{summary}

In summary, we have measured the lower critical field $H_{c1}$ in 
PrFeAsO$_{1-y}$ single crystals for {\boldmath $H$}$\parallel c$ and 
{\boldmath $H$}$\parallel ab$-plane by utilizing an array of miniature 
Hall sensor. Conventional methods using a single micro-Hall probe placed on the 
center of the crystal might overestimate $H_{c1}$ due to strong flux 
pinning. $H_{c1}$ measured by the sensor located very near to the edge of the crystal 
shows saturating behavior at low temperatures, which is consistent 
with the previous reports on the penetration depth measurements.
The anisotropy of $H_{c1}$ slightly decreases with decreasing temperature and is indicative of 
multiband superconductivity in PrFeAsO$_{1-y}$, in which the active band 
for superconductivity is more anisotropic.

\section*{ACKNOWLEDGEMENTS}
We thank A.~E.~Koshelev for useful discussion and T.~Terashima for 
technical assistance.
This work was supported by KAKENHI (No. 20224008) from JSPS, by Grant-in-Aid for 
the Global COE program ``The Next Generation of Physics, Spun from 
Universality and Emergence" and by Grant-in-Aid for Specially Promoted Research (No. 17001001) from MEXT, Japan.  
R.O. and H.S. were supported by the JSPS Research Fellowship for 
Young Scientists.\\

{\it Note added}: Recently, $H_{c1}$ measurements on NdFeAs(O,F) by using Hall probes are reported,\cite{Klein}
which show similar temperature dependence of $H_{c1}$.

\end{document}